\documentclass[%
 reprint,
 amsmath,amssymb,
 aps,
]{revtex4-2}

\usepackage{graphicx}
\usepackage{dcolumn}
\usepackage{bm}
\usepackage{braket}
\usepackage{mathtools}
\usepackage{here}
\usepackage{multirow}
\usepackage{color}
\usepackage[colorlinks=true, linkcolor=blue, citecolor=blue, urlcolor=blue]{hyperref}
\usepackage{natbib}
\begin{document}

\preprint{APS/123-QED}
\setlength\tabcolsep{5pt}

\title{Spin-$S$ Ising models with multispin interactions on the one-dimensional chain and two-dimensional square lattice}

\author{Kohei Suzuki}
\affiliation{Jij Inc., Bunkyo-ku, Tokyo 113-0031, Japan}

\begin{abstract}
We study spin-$S$ Ising models with $p$-spin interactions on the one-dimensional chain and the two-dimensional square lattice. 
Here, $S$ denotes the magnitude of the spin and $p$ represents the number of spins involved in each interaction.
The analysis is performed for $S=1/2,1,3/2,2$ and $p=3,4,5$.
For the one-dimensional model, we formulate transfer matrices and numerically diagonalize them to analyze the temperature dependence of the free energy and spin-spin correlations.
In the case of $S=1/2$, the free energy does not depend on $p$, and the spin-spin correlations are uniformly enhanced across all temperature scales as $p$ increases. 
In contrast, for $S \geq 1$, the free energy varies with $p$, and the spin-spin correlations are significantly enhanced at lower temperatures as $p$ increases.
For the two-dimensional model, by using multicanonical simulations, we analyze physical quantities such as an order parameter, internal energy, and specific heat. 
In addition, we define and examine an order parameter to distinguish ordered and disordered phases.
It is found that a first-order phase transition occurs at finite temperatures for all $S$ and $p\geq3$, and increasing $p$ strengthens its nature.
We present $S$ and $p$ dependence of the transition temperature and latent heat, and discuss effects of higher-order interactions on the nature of phase transitions.
\end{abstract}
\maketitle

\section{Introduction}
The Ising model, a representative example of classical spin systems, and its variations have been extensively studied for a long time since it was discussed by Lenz \cite{Lenz_1920} and Ising \cite{Ising_1925}.
In the field of statistical physics, the two-dimensional Ising model \cite{PhysRev.65.117} is one of the most well-known models to be analyzed exactly and exhibit a phase transition at finite temperatures.
In these models, two spins interact with each other through two-body interactions.
In terms of interactions, the Ising model can be extended by introducing $p$-body interactions with $p \geq 3$.
Systems with such higher-order interactions have been studied for a long time because of their intriguing properties.
For one-dimensional systems, although there are no phase transitions at finite temperatures, some studies have been conducted on the relation with satisfiability problems \cite{Fan_2011}, the free energy and spin correlations \cite{PhysRevB.27.2894,Turban_2016}, and the correspondence with two-dimensional systems under an external magnetic field \cite{Turban_2016}.

Regarding two-dimensional systems, they exhibit phase transitions at finite temperatures and have attracted significant interest owing to their remarkable features.
For instance, the eight-vertex model \cite{PhysRevB.2.723, PhysRevLett.26.832} is known to be mapped onto the square-lattice Ising model with two- and four-body interactions \cite{PhysRevB.4.2312, PhysRevB.4.3989}. 
On the triangular lattice, an Ising model with three-body interactions has been solved by Baxter and Wu \cite{PhysRevLett.31.1294, 1974AuJPh..27..357B, 1974AuJPh..27..369B}. 
While the transition temperature of this system is the same as that of the conventional square-lattice Ising model, its universality class corresponds to that of the four-state Potts model \cite{Potts_1952}.

Furthermore, an Ising model with $p$-body interactions in one direction and two-body interactions in the other on the rectangular lattice \cite{JPhysiqueLETTRES43295, MDebierre_1983} has been extensively studied.
Analysis utilizing self-duality reveals that the transition temperature is identical to that of the square-lattice Ising model, independent of $p$.
For $p=3$, the phase transition is second-order and its universality class belongs to that of four-state Potts model, and for $p \geq 4$, the transition becomes first-order.
As a further extension, an Ising model with $p$-body interactions in both directions on a square lattice is a natural extension of two-body interactions in the Ising model.
This model is also self-dual and the transition temperature is expected to be the same as that of the $p=2$ case \cite{LTurban_1982}.

All the studies mentioned so far have focused on $S=1/2$ spin systems.
One can extend the Ising model by considering general spin $S$. 
In general, such large spins can emerge as effective degrees of freedom through interactions.
From an application perspective, combinatorial optimization problems can be mapped onto Ising models \cite{fphy.2014.00005}, and spin-$S$ Ising models naturally correspond to integer programming problems.
In the case of usual two-body interactions, these types of models are known as Blume-Capel models \cite{PhysRev.141.517,CAPEL1966966}, which have been studied from the perspective of the tricritical point arising from the competition between uniaxial anisotropy and two-body interactions. 

So far we have mentioned two types of extension to the conventional Ising model: introducing $p$-spin  interactions and considering spin-$S$ systems.
While these extensions have been studied individually, there are few studies on models that incorporate both extensions.
The physical properties of such systems are quite interesting and these systems correspond to integer programming problems with higher-order terms in the context of mathematical optimization \cite{fphy.2014.00005}.
Thus, it is important to understand their basic properties for practical applications.

In this paper, we investigate two fundamental spin-$S$ Ising models with $p$-spin interactions, varying both $S$ and $p$ up to $S=2$ and $p=5$ to explore how these changes affect the properties of the system.
The first one is defined on a one-dimensional chain, where we can define transfer matrices and obtain numerically exact results.
The second one is on a two-dimensional square lattice, which shows phase transitions at finite temperatures and can be analyzed using classical Monte Carlo simulations.

This paper is organized as follows.
In Sec. \ref{sect-1dim}, we analyze the model on the one-dimensional chain. 
We define a transfer matrix and perform numerical analysis to investigate the free energy and spin-spin correlations. 
In Sec. \ref{sect-2dim}, we examine the two-dimensional square lattice model. 
We employ the multicanonical \cite{BERG1991249, PhysRevLett.68.9, PhysRevLett.69.2292} and the Wang-Landau methods \cite{PhysRevLett.86.2050, PhysRevE.64.056101}, examining the nature of phase transition through the typical physical quantities, such as an order parameter, internal energy, and specific heat. 
Finally, in Sec. \ref{sect-summary}, we summarize our results.

\section{One-Dimensional Systems}\label{sect-1dim}
It is well-known that one-dimensional classical spin systems with only finite-range interactions do not exhibit phase transitions at finite temperatures.
The model analyzed here does not show a phase transition either.
However, the behavior of certain physical quantities on $p$ remains interesting.
In this section, we analyze the one-dimensional spin-$S$ Ising model with $p$-spin interactions. 
We investigate the temperature dependence of the free energy and the spin correlation functions.
The former is a fundamental physical quantity because it contains all the thermodynamic information of the system, while the latter reflects the magnetic properties of the system.
By analyzing these quantities, we aim to clarify the effects of the number of spins involved in the interactions, $p$, on the system.

This section is organized as follows: 
First, we define the one-dimensional spin-$S$ Ising model with $p$-spin interactions. 
Next, we introduce the transfer matrix to calculate the free energy and spin correlation functions.
Finally, by numerically diagonalizing the transfer matrix, we examine the effects of $p$ on the free energy and on the correlation length of the spin correlation functions.

\subsection{Model}
\begin{figure}[!t]
    \centering
    \includegraphics[width=\columnwidth]{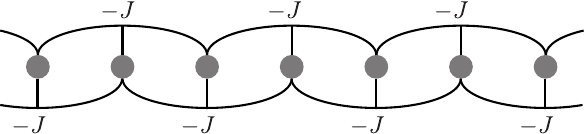}
    \caption{
      The schematic picture for the spin-$S$ Ising model with $p$-spin interactions on the one-dimensional chain for $p=3$.
      Gray circles represent the spins.
    }
    \label{1_dim_chain_p3}
\end{figure}

The one-dimensional spin-$S$ Ising model with $p$-spin interactions is defined by
\begin{align}
    &E_{\text{chain}}(\bm{s})=-J\sum^{N}_{i=1}\prod^{p-1}_{j=0}s_{i+j}.
    \label{ham_1_dim}
\end{align}
Here, $N$ represents the system size and $s_i$ denotes the spin variable at the site $i$.
We assume a ferromagnetic interaction $J > 0$, and impose the periodic boundary conditions: $s_{i+N}=s_i$ for $i=1,2,\ldots,N$.
The schematic illustration for the $p=3$ model is shown in Fig. \ref{1_dim_chain_p3}.
The values of $s_i$ are normalized by the magnitude of the spin $S$ as follows:
\begin{align}
    &s_{i} \in \Omega_{S},\quad\Omega_{S}= \left\{\frac{-S}{S},\frac{-S + 1}{S},\ldots,\frac{S}{S}\right\}.
    \label{spin_def}
\end{align}
$\Omega_{S}$ is the set of possible values for spin variables. 
For example, $\Omega_{1/2}=\{-1, +1\}$ for $S=1/2$, and $\Omega_{1}=\{-1, 0, +1\}$ for $S=1$.
This normalization restricts the values of the spins to $-1 \leq s_i \leq +1$, which leads to $-JN \leq E_{\text{chain}}(\bm{s}) \leq JN$ ensuring comparable energy scales for different $p$ and $S$.
To simplify some of the calculations, we restrict $N$ to be a multiple of $p$, i.e., it is assumed that $N=pM$ with $M$ being a non-negative integer, which does not affect the results in the thermodynamic limit ($N\rightarrow\infty$).

\subsection{Transfer matrix}
Next, we introduce the transfer matrix to calculate the partition function.
Let us start by rewriting Eq. (\ref{ham_1_dim}) using the assumption that $N$ is a multiple of $p$: $N=pM$.
By dividing Eq. (\ref{ham_1_dim}) into terms for each $p$, we get
\begin{align}
    E_{\text{chain}}(\bm{s})=-J\sum^{M}_{i=1}\sum^{p}_{k=1}\prod^{k-1}_{j=-p+k}s_{pi+j}.
\end{align}
This allows us to express the partition function as
\begin{align}
    Z
    &=\sum_{\bm{s}}\exp\left[-\beta E_{\text{chain}}(\bm{s})\right]\nonumber\\
    &=\sum_{\bm{s}}\prod^{M}_{i=1}\exp\left(\beta J\sum^{p}_{k=1}\prod^{k-1}_{j=-p+k}s_{pi+j}\right),
\end{align} 
with $\beta=1/T$ being inverse temperature.
Here $\sum_{\bm{s}}=\sum_{s_1}\sum_{s_2},\ldots,\sum_{s_N}$ represents the summation over all spin variables.
One can carry out the summation over the spin variables with indices that are multiples of $p$:
\begin{align}
    Z
    &=\sum_{\bm{s}\setminus\bm{s}_{p}}\sum_{\bm{s}_p}\prod^{M}_{i=1}\exp\left(\beta J\sum^{p}_{k=1}\prod^{k-1}_{j=-p+k}s_{pi+j}\right)\nonumber\\
    &=\sum_{\bm{s}\setminus\bm{s}_{p}}\prod^{M}_{i=1}\sum_{s_{pi}\in\Omega_{S}}\exp\left(\beta J\sum^{p}_{k=1}\prod^{k-1}_{j=-p+k}s_{pi+j}\right),
\end{align}
where $\sum_{\bm{s}_{p}}=\sum_{s_p}\sum_{s_{2p}}\cdots\sum_{s_{Mp}}$ represents the summation over the spin variables with indices that are multiples of $p$, and $\sum_{\bm{s}\setminus\bm{s}_{p}}$ represents the summation over the remaining spin variables.
The term $\sum_{s_{pi}\in\Omega_{S}}\exp(\cdots)$ depends on the $2p-2$ spin variables: $s_{pi-p+1},s_{pi-p},\ldots,s_{pi-1}$ and $s_{pi+1},s_{pi+2},\ldots,s_{pi+p-1}$.
This term can be considered as a matrix element of the transfer matrix $\hat{T}$.
We introduce its element as 
\begin{align}
    &\hat{T}_{(s_{pi-p+1},\ldots,s_{pi-1}), (s_{pi+1},\ldots,s_{pi+p-1})}\nonumber\\
    &=\sum_{s_{pi}\in\Omega_{S}}\exp\left(\beta J\sum^{p}_{k=1}\prod^{k-1}_{j=-p+k}s_{pi+j}\right).
    \label{transfer_matrix}
\end{align}
The partition function can be expressed in terms of this transfer matrix as follows:
\begin{align}
    Z
    &=\sum_{\bm{s}\setminus\bm{s}_{p}}\prod^{M}_{i=1}\hat{T}_{(s_{pi-p+1},\ldots,s_{pi-1}), (s_{pi+1},\ldots,s_{pi+p-1})}\nonumber\\
    &=\sum_{s_1}\cdots\sum_{s_{p-1}}\hat{T}^{M}_{(s_1,\ldots,s_{p-1}), (s_{pM+1},\ldots,s_{pM+p-1})}\nonumber\\
    &=\sum_{s_1}\cdots\sum_{s_{p-1}}\hat{T}^{M}_{(s_1,\ldots,s_{p-1}), (s_1,\ldots,s_{p-1})}\nonumber\\
    &=\text{Tr}[\hat{T}^M].
\end{align}
Note that we use the relations $s_{i+N}=s_i$ for $i=1,2,\ldots,N$ and $N=pM$.
As an example, we present the transfer matrix for $S=1/2$ and $p=2$.
The matrix elements are given by
\begin{align}
    &\hat{T}_{(s_{2i-1}),(s_{2i+1})}\nonumber\\
    &=\sum_{s_{2i}\in\{-1,+1\}}\exp\left(\beta J(s_{2i-1}s_{2i} + s_{2i}s_{2i+1})\right)\nonumber\\
    &=2\cosh(\beta J (s_{2i-1} + s_{2i+1})),
\end{align}
and the transfer matrix is
\begin{align}
    \hat{T}&=
    \begin{pmatrix}
    \hat{T}_{(+1),(+1)} & \hat{T}_{(+1),(-1)} \\
    \hat{T}_{(-1),(+1)} & \hat{T}_{(-1),(-1)} \\
    \end{pmatrix}\nonumber\\
    &=2\begin{pmatrix}
    \cosh(2\beta J) & 1 \\
    1 & \cosh(2\beta J) \\
    \end{pmatrix}.
\end{align}
The transfer matrix introduced here is a real square matrix of size $(2S+1)^{(p-1)}\times (2S+1)^{(p-1)}$.
Denoting its eigenvalues by $\lambda_1 \geq \lambda_2 \geq \cdots \lambda_{(2S+1)^{p-1}}$, the partition function can finally be expressed as:
\begin{align}
    Z=\text{Tr}\left[\hat{T}^M\right]=\sum^{(2S+1)^{p-1}}_{i=1}\lambda^{M}_i.
\end{align}

\subsection{Free energy}
Now, we calculate the free energy and examine its dependence on $p$. 
The free energy per spin can be obtained from the partition function as follows:
\begin{align}
    f_{p}(T)
    &=-T\lim_{N\rightarrow\infty}\frac{1}{N}\log Z\nonumber\\
    &=-T\lim_{N\rightarrow\infty}\frac{1}{N}\log \lambda^{M}_1\nonumber\\
    &=-T\log\lambda^{\frac{1}{p}}_1,
\end{align}
where $\log Z$ represents the natural logarithm of $Z$ and $T$ is the temperature. 
Note that we set the Boltzmann constant $k_{\text{B}}$ as $1$.
The free energy can be determined by finding the largest eigenvalue $\lambda_1$ of the transfer matrix $\hat{T}$.
For $S=1/2$, the largest eigenvalue for any $p$ can be calculated analytically \cite{Turban_2016} as
\begin{align}
    \lambda_1=\left[2\cosh(\beta J)\right]^{p}.
\end{align}
Thus, the free energy for $S=1/2$ is
\begin{align}
    f_p(T)=-T\log\left[2\cosh\left(\beta J\right)\right],
\end{align}
which is equal to that of the conventional one-dimensional Ising model and independent of $p$.

\begin{figure}[!t]
  \centering
  \includegraphics[width=\columnwidth]{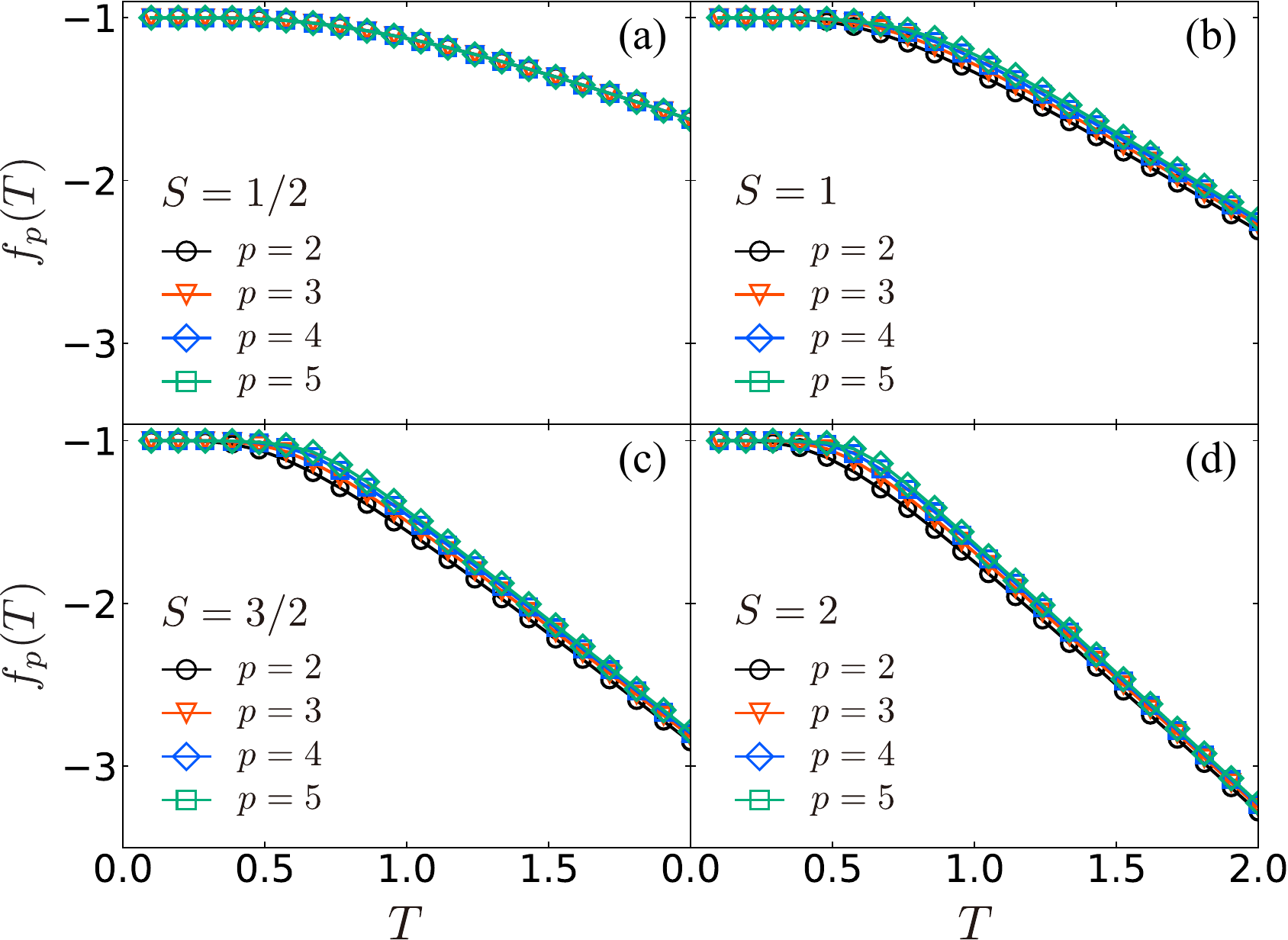}
  \caption{
    Temperature $T$ dependence of the free energy per site $f_p(T)$ for (a) $S=1/2$, (b) $S=1$, (c) $S=3/2$, and (d) $S=2$.
    In the case of $S=1/2$, the free energy takes the form $f_p(T) = -T\log\left[2\cosh\left(\beta J\right)\right]$, which is independent of $p$.
  }
  \label{1_dim_free_energy}
\end{figure}

In contrast, for $S \geq 1$, the free energy depends on $p$. 
Since it is difficult to calculate analytically for $S \geq 1$, we confirm this by numerically diagonalizing the transfer matrix. 
The results are shown in Fig. \ref{1_dim_free_energy}.
For $S=1/2$, one can see that the free energy actually does not change with $p$ [Fig. \ref{1_dim_free_energy}(a)],
while for $S \geq 1$, the free energy increases as $p$ increases [Figs. \ref{1_dim_free_energy}(b)-\ref{1_dim_free_energy}(d)] especially for the intermediate temperature range. 
For low temperatures, the influence of $p$ is small, and the free energy converges to $-1$, reflecting the fact that the ground state energy per site is $-1$ for any $p$.

In the case of $S=1/2$, since the product of any number of spins only takes the values $-1$ or $+1$, eliminating $p$ dependence, this product can be regarded as a new spin variable: $\prod_{i}s_i \in \{-1, +1\}$. 
As a result, the free energy becomes independent of $p$.
By using this property, one can directly calculate the free energy for $S=1/2$ without defining the transfer matrix.
For $S \geq 1$, this property does not hold and the free energy depends on $p$.

\subsection{Spin-spin correlations}
\begin{figure}[!t]
    \centering
    \includegraphics[width=\columnwidth]{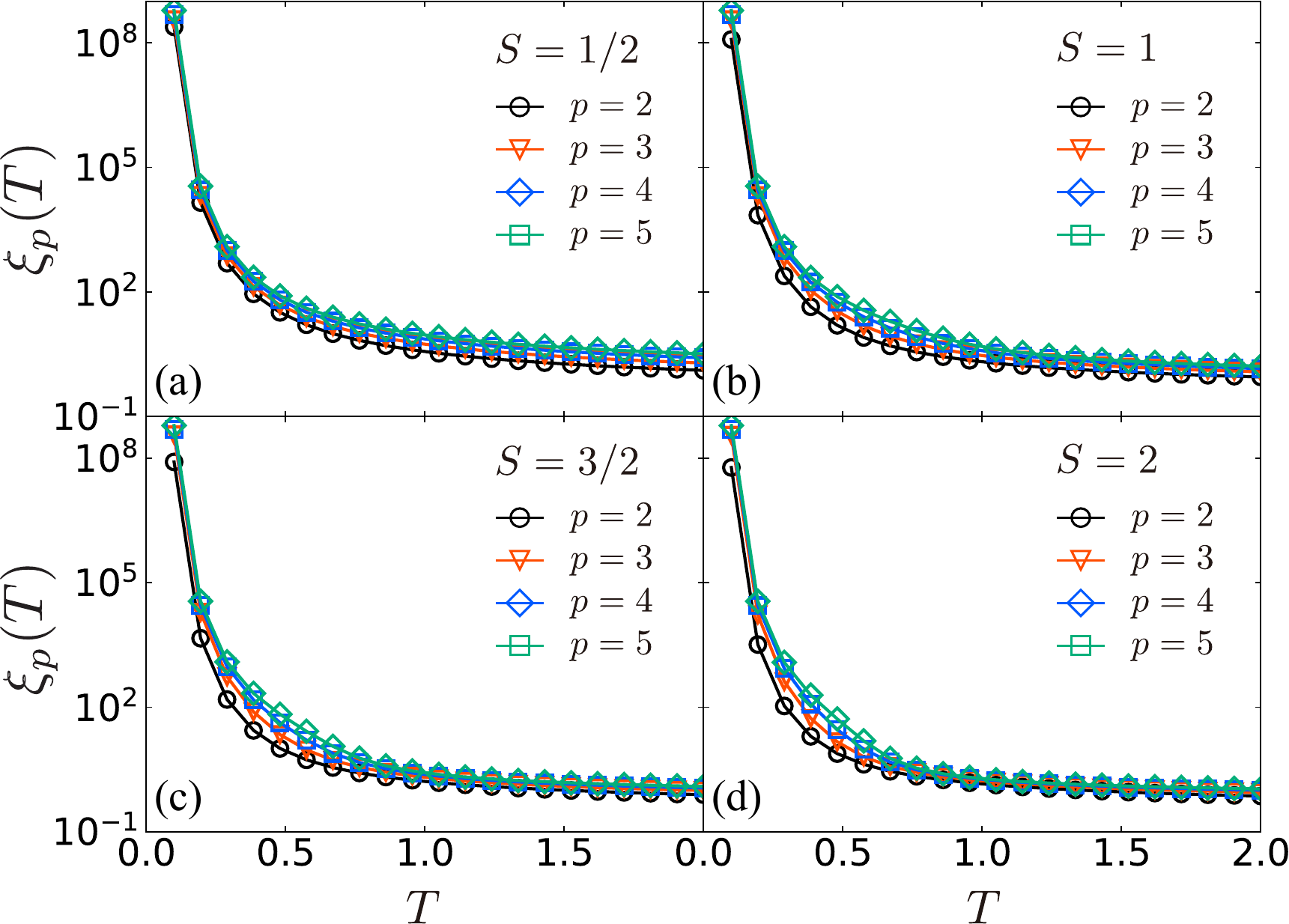}
    \caption{
      Temperature $T$ dependence of the spin-spin correlation length $\xi_p(T)$ for (a) $S=1/2$, (b) $S=1$, (c) $S=3/2$, and (d) $S=2$.
      In the case of $S=1/2$, the correlation length takes the form $\xi_p(T)=-p/[2\log\tanh(\beta J)]$.
    }
    \label{1_dim_correlation_length}
\end{figure}
The spin correlation function reflects the magnetic properties of the system. 
Since the model analyzed here does not show a phase transition, the spin correlation function exhibits exponential decay at finite temperatures for any $S$ and $p$. 
However, by comparing the correlation length, we can investigate the effects of the number of spins involved in the interactions, $p$, on the magnetic properties of the system.
The spin correlation function is defined as follows:
\begin{align}
    \braket{s_i s_j} = \frac{1}{Z}\sum_{\bm{s}} s_i s_j \exp\left[-\beta E_{\text{chain}}(\bm{s})\right].
\end{align}
From the symmetry of the system discussed in Ref. [\citealp{Turban_2016}], $\braket{s_i s_j}$ for any $S$ and $p$ takes zero if the distance between two spins is not a multiple of $p$:
\begin{align}
    \braket{s_i s_j} = 0, \quad |j - i| \neq mp,
\end{align}
where $m$ is a non-negative integer. 
On the other hand, if the distance between the spins is a multiple of $p$, $\braket{s_i s_j}$ decays exponentially as follows:
\begin{align}
    \braket{s_i s_j} = a \exp\left(-\frac{|j - i|}{\xi_{p}(T)}\right), \quad |j - i| = mp.
    \label{spin_cf_1}
\end{align}
Here, $a$ only depends on the temperature, and $\xi_{p}(T)$ is the correlation length. 
It is well known that the correlation length can be expressed using the largest and second largest eigenvalues of the transfer matrix. 
Since the transfer matrix in Eq. (\ref{transfer_matrix}) connects spins separated by a distance of $p$, the spin-spin correlation function takes the following form:
\begin{align}
    \braket{s_i s_j} = a \left(\frac{\lambda_2}{\lambda_1}\right)^\frac{|i-j|}{p}.
    \label{spin_cf_2}
\end{align}
From Eqs. (\ref{spin_cf_1}) and (\ref{spin_cf_2}), the correlation length can be obtained using the largest eigenvalue $\lambda_1$ and the second largest eigenvalue $\lambda_2$:
\begin{align}
    \xi_{p}(T) = -\frac{p}{\log\frac{\lambda_2}{\lambda_1}}.
\end{align}

\begin{figure}[!t]
    \centering
    \includegraphics[width=\columnwidth]{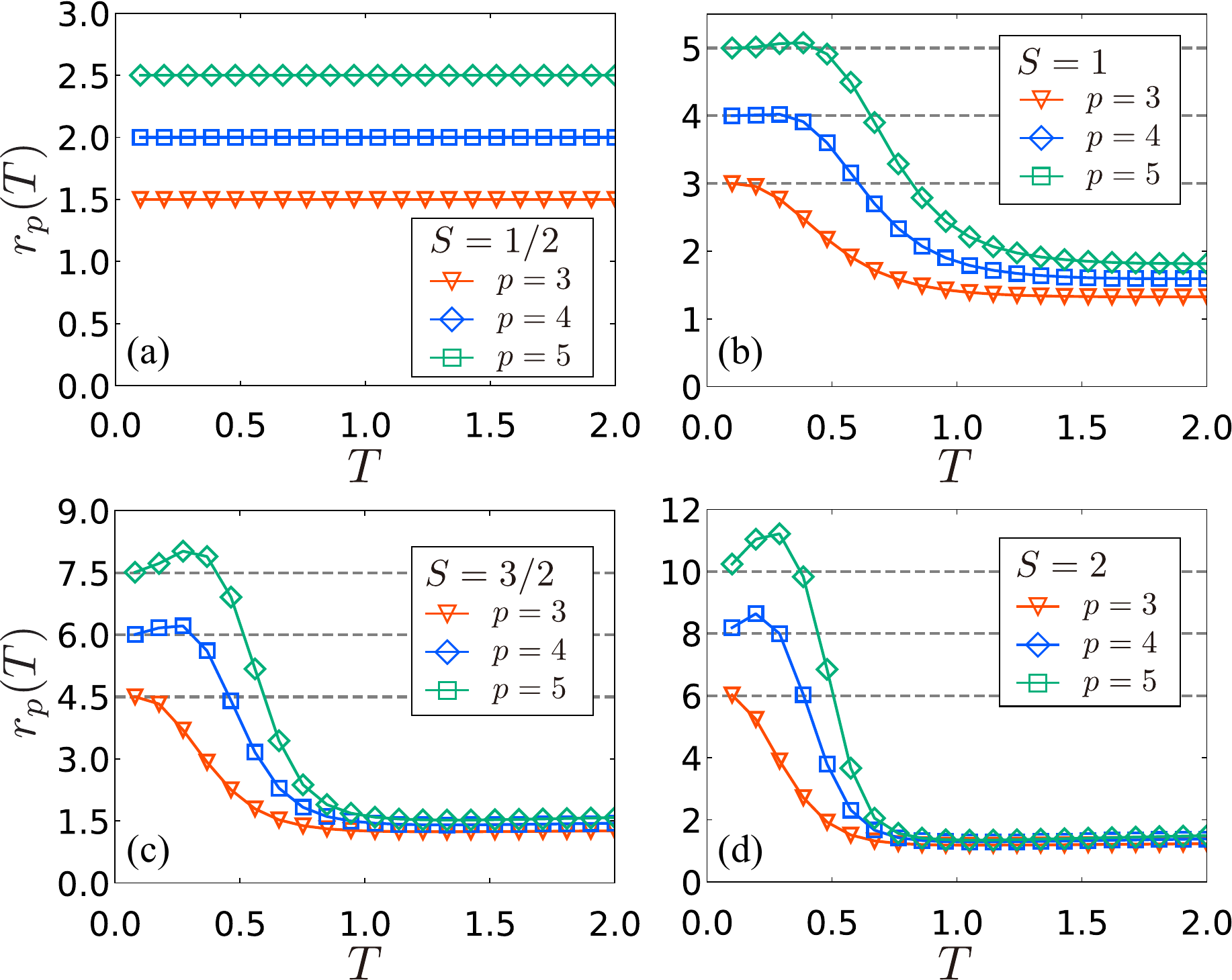}
    \caption{
      Temperature $T$ dependence of the ratio of correlation lengths $r_p(T)$ for (a) $S=1/2$, (b) $S=1$, (c) $S=3/2$, and (d) $S=2$.
      In the case of $S=1/2$, the ratio takes the form $r_p(T)=p/2$.
      The dashed lines indicate the value corresponding to $Sp$.
    }
    \label{1_dim_correlation_length_ratio}
\end{figure}

Now, let us discuss the results.
Figure \ref{1_dim_correlation_length} shows the temperature dependence of $\xi_{p}(T)$.
One can see that the correlation length increases for any $S$ as $p$ increases.
This is because increasing $p$ extends the range of the interaction, making it easier for spins to align. 
To make it easier to see the effects of adjusting $p$, we define the ratio of correlation lengths as
\begin{align}
    r_p(T)=\frac{\xi_{p}(T)}{\xi_{p=2}(T)}.
\end{align}
This ratio represents the magnification of $\xi_{p}(T)$ compared to the correlation length for $p=2$. 
The temperature dependence of $r_p(T)$ is shown in Fig. \ref{1_dim_correlation_length_ratio}.
For $S=1/2$, $r_p$ does not depend on the temperature and takes a constant value: $r_p = Sp = \frac{p}{2}$. 
This can be analytically calculated and the correlation length is given by $\xi_p = -p/[2\log\tanh(\beta J)]$ \cite{Turban_2016}, confirming that $r_p = p/2$. 
In contrast, for $S \geq 1$, $r_p$ depends on the temperature.
However, at low temperatures, $r_p$ shows a similar behavior and converges to $r_p = Sp$ at zero temperature. 
Furthermore, for larger $S$ and $p$, $r_p$ reaches the maximum value at low temperatures.
We conclude that the number of spins involved in the interaction $p$ enhances the magnetic properties of the system mostly in the low temperature region and the correlation length takes the simple form $\xi_{p}(T)\simeq Sp\times\xi_{p=2}(T)$ around zero temperature.

\section{Two-dimensional Systems}\label{sect-2dim}
Unlike the one-dimensional model analyzed previously [Eq. (\ref{ham_1_dim})], 
two-dimensional classical spin systems exhibit phase transitions at finite temperatures.
In this section, we explore how the number of interacting spins, $p$, influences the nature of phase transitions.
In the following, we first introduce the spin-$S$ Ising model with $p$-spin interactions on the two-dimensional square lattice.
We analyze the models for $S=1/2,1,3/2,3$ and $p=3,4,5$ as in the one-dimensional models.
Next, an order parameter is introduced to distinguish between the ordered and disordered phases. 
We then briefly explain the numerical methods used to analyze the model.
Finally, we discuss the results and the nature of the phase transitions.

\subsection{Model and order parameter}
We start by introducing the spin-$S$ Ising model with $p$-spin interactions on the two-dimensional square lattice.
The model is described by
\begin{align}
    E_{\text{sq}}(\bm{s})=-J\sum^{L}_{i_x=1}\sum^{L}_{i_y=1}\left(\prod^{p-1}_{k=0}s_{i_x,i_y+k}+\prod^{p-1}_{k=0}s_{i_x+k,i_y}\right).
    \label{ham_2_dim}
\end{align}
Here, $L$ represents the length of a side of the square lattice, and $p \geq 2$ is the number of interacting spins.
To suppress finite size effects, we restrict $L$ to be a multiple of $p$, i.e. $L=pM$ with $M$ being a non-negative integer.
This restriction does not affect thermodynamic-limit results and simplifies their estimation.
$s_{i_x,i_y}$ denotes the spin variable at the coordinate $(i_x, i_y)$ and these spins are normalized to take values in $\Omega_{S}$. See Eq. (\ref{spin_def}) for details.
The periodic boundary conditions are imposed to the system and the spin variables satisfy $s_{i_x+L,i_y} = s_{i_x,i_y+L} = s_{i_x,i_y}$ for $i_x,i_y=1,2,\ldots,L$. 
$J>0$ is the magnitude of the interactions and we set $J=1$ as an energy unit.
Under these conditions, the energy range of $E_{\text{sq}}(\bm{s})$ for any $p$ and $S$ is $-2L^2 \leq E_{\text{sq}}(\bm{s}) \leq 2L^2$ and the energy scales are comparable for different $p$ and $S$.
Note that $E_{\text{sq}}(\bm{s})$ for $p=2$ and $S=1/2$ corresponds to the conventional Ising model on the square lattice.

Next, we define an order parameter to determine whether the system is in an ordered phase or not. 
The magnetization $m=\sum^{L}_{i_x=1} \sum^{L}_{i_y=1} s_{i_x,i_y}/L^2$ is commonly used to distinguish the magnetic phases for conventional Ising models.
Since the spin-inversion symmetry does not break and the magnetization $m$ takes always zero for the finite size systems, the squared magnetization $m^2$ is also used for finite-size numerical calculations.

\begin{figure}[!t]
  \centering
  \includegraphics[width=\columnwidth]{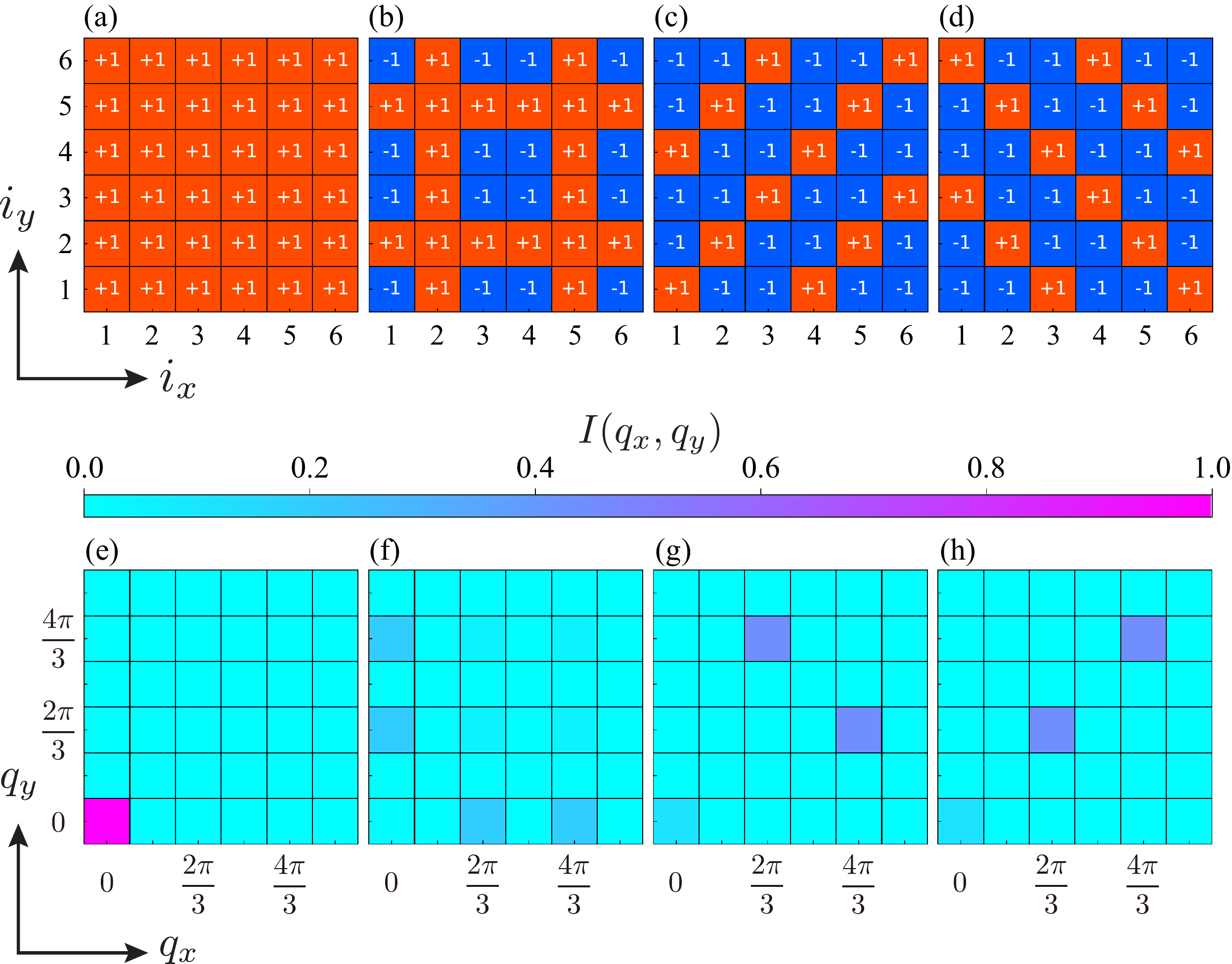}
  \caption{
    Typical ground-state spin configurations (a)--(d) and their Fourier intensities $I(q_x,q_y)$ (e)--(h) for $S=1/2$, $p=3$, and $L=6$.
    The configurations and their Fourier intensities correspond to (a) and (e), (b) and (f), (c) and (g), and (d) and (h), respectively.
  }
  \label{2_dim_p3_ground_state}
\end{figure}

For our models, it is true that the expectation value of the squared magnetization $\braket{m^2}$ takes non-zero values in the ordered phase.
Let us confirm this for the $p=3$ model as an example.
Considering the ground states, the spin configurations for $p=3$ are 16-fold degenerate.
In general, the ground state is $2^{(p-1)^2}$-fold degenerate if $L$ is a multiple of $p$.
We show typical ground states for $L=6$ in Figs. \ref{2_dim_p3_ground_state}(a)-\ref{2_dim_p3_ground_state}(d).
In addition, there are eight similar ground states obtained by using translational symmetry from the configuration in Fig. \ref{2_dim_p3_ground_state}(b) and two similar ground states each from Figs. \ref{2_dim_p3_ground_state}(c) and \ref{2_dim_p3_ground_state}(d), resulting in a total of 16 states.
One can easily confirm that the squared magnetization for each state is $\braket{m^2}=1$ for Fig. \ref{2_dim_p3_ground_state}(a), $\braket{m^2}=1/81$ for Fig. \ref{2_dim_p3_ground_state}(b), and $\braket{m^2}=1/9$ for Figs. \ref{2_dim_p3_ground_state}(c) and \ref{2_dim_p3_ground_state}(d).
Since each of the 16 states appears randomly in the ground state, the expectation value of the squared magnetization is calculated as $\braket{m^2}=(1 + 9/81 + 6/9)/16=1/9$, and from similar calculations, one can obtain $\braket{m}=0$ in the ground state.
Although the squared magnetization $m^2$ takes nonzero values at low temperatures, there are more relevant order parameters for the systems with $p \geq 3$.

To define more relevant order parameters, we here consider the Fourier intensity:
\begin{align}
    I(q_x, q_y) = \left| \frac{1}{L^2} \sum^{L}_{i_x=1} \sum^{L}_{i_y=1} s_{i_x,i_y} e^{i(q_xi_x + q_yi_y)} \right|^2.
\end{align}
Here, $q_x,q_y=0,2\pi/L,\ldots,2\pi(L-1)/L$ is the wave number and $i$ represents the imaginary unit.
The squared magnetization corresponds to $I(0, 0)$ and the Fourier intensity $I(q_x, q_y)$ introduced here can be regarded as a generalization of the order parameter for classical spin systems.
Figures \ref{2_dim_p3_ground_state}(e)-\ref{2_dim_p3_ground_state}(h) show $I(q_x,q_y)$ for the configurations shown in Figs. \ref{2_dim_p3_ground_state}(a)-\ref{2_dim_p3_ground_state}(d).
The configurations and their Fourier intensities correspond to Figs. \ref{2_dim_p3_ground_state}(a) and \ref{2_dim_p3_ground_state}(e), \ref{2_dim_p3_ground_state}(b) and \ref{2_dim_p3_ground_state}(f), \ref{2_dim_p3_ground_state}(c) and \ref{2_dim_p3_ground_state}(g), and \ref{2_dim_p3_ground_state}(d) and \ref{2_dim_p3_ground_state}(h) respectively.
Although there are 16 states in the ground state, the Fourier intensities are translation invariant and there are only four types as shown in Figs. \ref{2_dim_p3_ground_state}(e)-\ref{2_dim_p3_ground_state}(h).
One can see that $I(q_x,q_y)$ takes nonzero values in the case that $(q_x,q_y) \in \{0,2\pi/3, 4\pi/3\}\times\{0,2\pi/3, 4\pi/3\}$.
The sum of these Fourier intensities takes 1 in the ground state for $p=3$.
Generalizing these arguments, we define the order parameter for $p\geq 3$ as
\begin{align}
    O = \sum_{q_x \in Q_p} \sum_{q_y \in Q_p} I(q_x, q_y),
    \label{def_order_params}
\end{align}
where
\begin{align}
    Q_p = \left\{\frac{2n\pi}{p} \middle| n = 0, 1, \ldots, p-1\right\}.
\end{align}
One can easily calculate the canonical expectation values of $O$ [Eq. (\ref{def_exp_O})] in the ground state and high temperature limit:
\begin{align}
    \lim_{T\rightarrow 0} \braket{O}=1,\quad
    \lim_{T\rightarrow\infty} \braket{O}=\frac{p^2}{L^2}.
\end{align}
Note that in the thermodynamic limit ($L\rightarrow \infty$), $\braket{O}$ becomes zero at high temperatures. 
We use the order parameter $O$ defined in Eq. (\ref{def_order_params}) to distinguish the ordered and disordered phases. $\braket{O}\sim 1$ at finite temperatures indicates that the system is in the ordered phase.

\subsection{Methods}
As will be discussed later, the model analyzed here exhibits a first-order phase transition.
Conventional classical Monte Carlo simulations such as the Metropolis method \cite{METROPOLIS,METROPOLIS_HASTING} and the heat bath method \cite{HEATBATH} often struggle with such transitions since the systems get trapped in metastable states which is typical of first-order phase transitions.
Cluster update methods such as the Swendsen-Wang algorithm \cite{PhysRevLett.58.86} and Wolff algorithm \cite{PhysRevLett.62.361} are known to be powerful computational methods for the standard two-dimensional Ising models.
These techniques overcome critical slowing down associated with second-order phase transitions.
However, it is difficult to directly apply these methods to systems with higher-order interactions.
One method that has overcome these difficulties is the self-learning Monte Carlo method \cite{PhysRevB.95.041101}, which has indeed been applied to Ising models with higher-order interactions, in which the system exhibits second-order phase transitions.

In this paper, we employ the multicanonical method \cite{BERG1991249, PhysRevLett.68.9, PhysRevLett.69.2292} and the Wang-Landau method \cite{PhysRevLett.86.2050, PhysRevE.64.056101}, which are particularly effective for systems undergoing first-order phase transitions.
These methods sample states by performing a random walk in the energy space, effectively avoiding traps in metastable states.
Our analysis proceeded as follows.
Initially, we utilized the Wang-Landau method to estimate the density of states $D(E)$ with $E$ being the energy of the system.
We updated the modification factor $f$ (see Ref. [\citealp{PhysRevLett.86.2050}] for details) to $\sqrt{f}$ when the minimum value of the energy histogram reached 95\% of its average, continuing until $f$ reached $e^{10^{-8}}\simeq 1+10^{-8}$.
The energy range was divided into up to 16 segments, and Wang-Landau calculations for each energy segment were carried out in parallel. 
Single-threaded calculations were also performed using the symmetry of the system without dividing the energy range.
In this case, all negative energies that appeared during the simulation were treated as positive energies and occurrences of the state with $E=0$ were counted twice.
From these treatments we have obtained $D(E)$ with $E \geq 0$ approximately twice as fast.
Finally, the density of states in the negative energy range was determined so that $D(-E)=D(E)$. 
This method has the advantage that the boundary effects do not appear near $E=0$, compared to the method in which the simulation is limited to the region with positive or negative energy range.
Note, however, that this method loses its advantages if parallel calculations are performed by dividing the energy range. 
If the energy range is divided, the boundary effects may appear at the joints of the energy segments.

After estimating $D(E)$, we performed the multicanonical simulations and improved the estimated $D(E)$.
During each sweep of the simulations, we also stored the order parameter $O(E)$ [Eq. (\ref{def_order_params})] every time a state was updated.
To reduce computation time, instead of recalculating the order parameter every time, the order parameter is calculated by updating its difference.
The number of sweeps conducted was $10^8$ for the models with $S=1/2$ and $1$, and $10^9$ for those with $S=3/2$ and $2$.
All calculations were performed five times for system sizes up to $L^2=60\times 60$.

\subsection{Results}
In this subsection, we explain simulation results for the spin-$S$ Ising model with $p$-body interactions on a square lattice [Eq. (\ref{ham_2_dim})] with $S=1/2,1,3/2,2$ and $p=3,4,5$. 
First, the temperature dependence of the order parameter [Eq. (\ref{def_order_params})] and internal energy will be shown.
Next, by examining the energy distribution near the transition point, we confirm that the system shows a first-order phase transition. 
We then extrapolate the latent heat and transition temperature associated with the first-order transition in the thermodynamic limit. 
Finally, to verify the accuracy of the transition temperature by a different method, we calculate the Binder ratio of the order parameter for $p=3$ by the conventional Monte Carlo simulations with single spin flips.

\begin{figure}[t]
  \centering
  \includegraphics[width=\columnwidth]{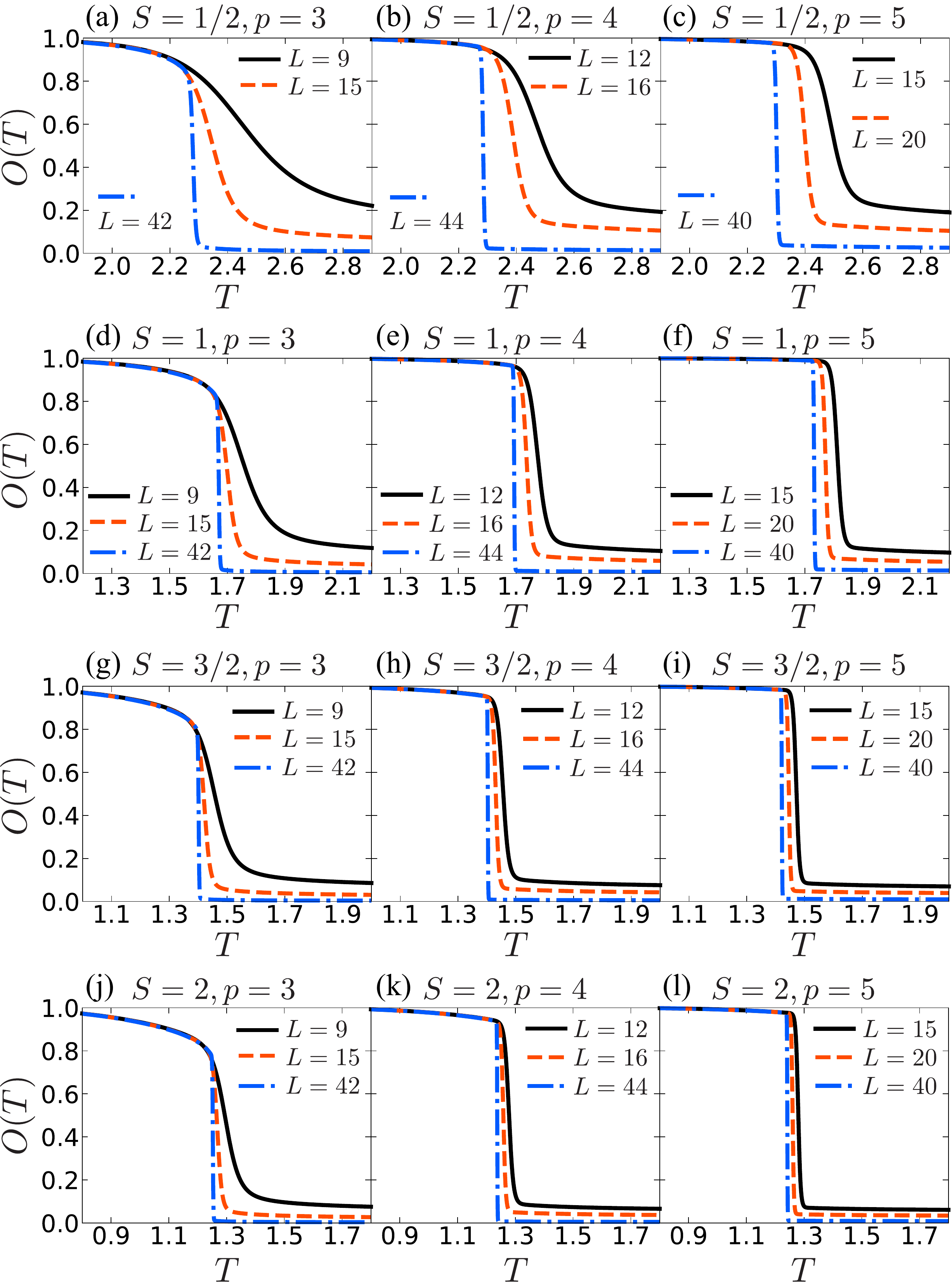}
  \caption{
    Temperature $T$ dependence of the order parameter $O(T)$ for
    (a) $S=1/2,p=3$, (b) $S=1/2,p=4$, (c) $S=1/2,p=5$, (d) $S=1,p=3$, (e) $S=1,p=4$, (f) $S=1,p=5$,
    (g) $S=3/2,p=3$, (h) $S=3/2,p=4$, (i) $S=3/2,p=5$, (j) $S=2,p=3$, (k) $S=2,p=4$, and (l) $S=2,p=5$.
    Average values from five independent runs are shown here.
    The maximum standard deviation is $\sim 10^{-5}$ and too small to be visible on the graph for all results.
  }
  \label{2_dim_op_temp_deps}
\end{figure}

We begin by presenting the results for the order parameter $O(T)$, which is defined as the expectation value of $O$ [Eq. (\ref{def_order_params})]:
\begin{align}
    O(T)\coloneqq \braket{O}=
    \frac{\sum_{E}D(E)O(E)e^{-\beta E}}{\sum_{E}D(E)e^{-\beta E}}.
    \label{def_exp_O}
\end{align}
Here, $O(E)$ is the order parameter corresponding to the energy $E$ and obtained through multicanonical simulations.
$D(E)$ is the density of states and $\beta=1/T$ is the inverse temperature.
We show temperature dependence of $O(T)$ for $S=1/2,1,3/2,2$ and $p=3,4,5$ in Fig. \ref{2_dim_op_temp_deps}.
One can see that for all $p$ and $S$, the order parameter $O(T)$ approaches zero at high temperatures and 1 at low temperatures.
Additionally, in the intermediate temperature range, $O(T)$ changes abruptly from zero to a positive value. 
These tendencies become more pronounced as the system size $L^2$ increases.
In the thermodynamic limit, $O(T)$ is expected to jump from zero to a positive value, suggesting that the system undergoes a first-order phase transition.
One can also see that the jump in the order parameter at the transition point tends to increase with the rise in $S$ and $p$.

The estimated transition temperatures for $p=3,4,5$ are $T_{\text{c}}\sim 2.3$ for $S=1/2$, $T_{\text{c}}\sim 1.6$ for $S=1$, $T_{\text{c}}\sim 1.4$ for $S=3/2$, and $T_{\text{c}}\sim 1.2$ for $S=2$.
The transition temperature decreases as the magnitude of the spin increases.
This tendency aligns with the behavior for $p=2$ models, where the temperature of second-order phase transition is $T_c=2.269...$ \cite{PhysRev.65.117} for $S=1/2$, $T_c\simeq 1.68-1.71$ \cite{PhysRevB.33.1717,PhysRevB.57.11575,PhysRevE.73.036702,PhysRevE.81.041113,BUTERA201822} for $S=1$, $T_c\simeq 1.46$ \cite{PhysRevB.57.11575} for $S=3/2$, and $T_c\simeq 1.32-1.69$ \cite{TUCKER198727,KANEYOSHI1992495,JURCISIN1996684,HACHEM2017927} for $S=2$, indicating a consistent decrease as $S$ increases.

\begin{figure}[t]
  \centering
  \includegraphics[width=\columnwidth]{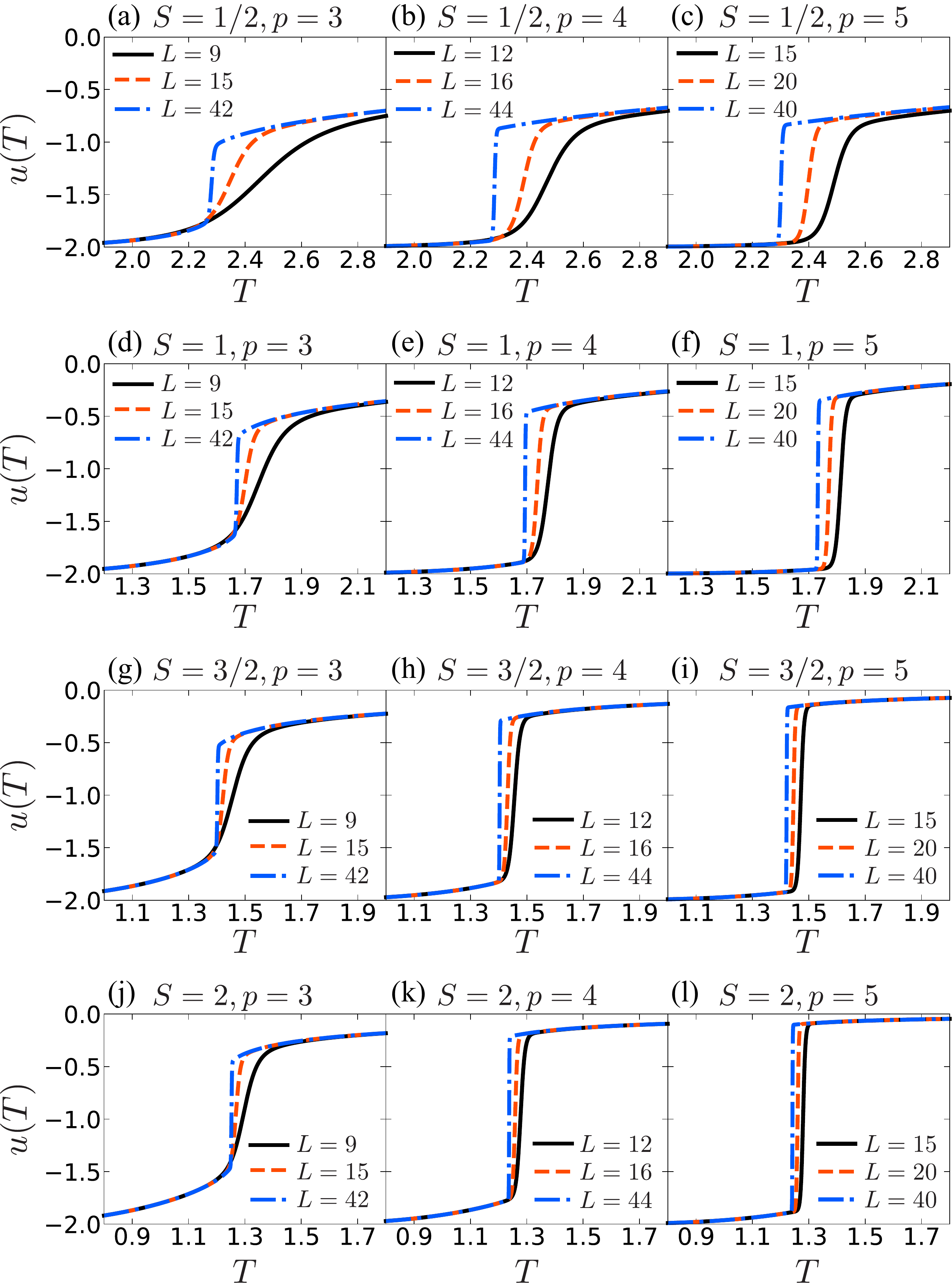}
  \caption{
    Temperature $T$ dependence of the internal energy $u(T)$ for
    (a) $S=1/2,p=3$, (b) $S=1/2,p=4$, (c) $S=1/2,p=5$, (d) $S=1,p=3$, (e) $S=1,p=4$, (f) $S=1,p=5$,
    (g) $S=3/2,p=3$, (h) $S=3/2,p=4$, (i) $S=3/2,p=5$, (j) $S=2,p=3$, (k) $S=2,p=4$, and (l) $S=2,p=5$.
    Average values from five independent runs are shown here.
    The maximum standard deviation is $\sim 10^{-4}$ and too small to be visible on the graph for all results.
  }
  \label{2_dim_energy_temp_deps}
\end{figure}

Next, we show results of the internal energy density defined by
\begin{align}
    u(T)\coloneqq \frac{\braket{E}}{L^2}= \frac{1}{L^2}\frac{\sum_{E}D(E)Ee^{-\beta E}}{\sum_{E}D(E)e^{-\beta E}}.
\end{align}
Since the spin variables are normalized to take values from $-1$ to $1$ [Eq. (\ref{spin_def})], the energy density in the ground state is $-2$ for all $p$ and $S$.
The temperature dependence of the internal energy $u(T)$ for $S=1/2,1,3/2,2$ and $p=3,4,5$ is shown in Fig. \ref{2_dim_energy_temp_deps}, and indeed, it converges to $-2$ at low temperatures.
The important thing is that $u(T)$ also changes abruptly at finite temperatures, and this change occurs at the same temperature ranges predicted by the order parameter as shown in Fig. \ref{2_dim_op_temp_deps}. 
Additionally, the abrupt changes in $u(T)$ become more pronounced as the system size increases, and its change increases with $p$ and $S$.
These results indicate that the system exhibits the first-order phase transition, consistent with the results from the order parameter. 
The results also suggest that increasing $p$ and $S$ results in larger latent heat.

\begin{figure}[!t]
  \centering
  \includegraphics[width=\columnwidth]{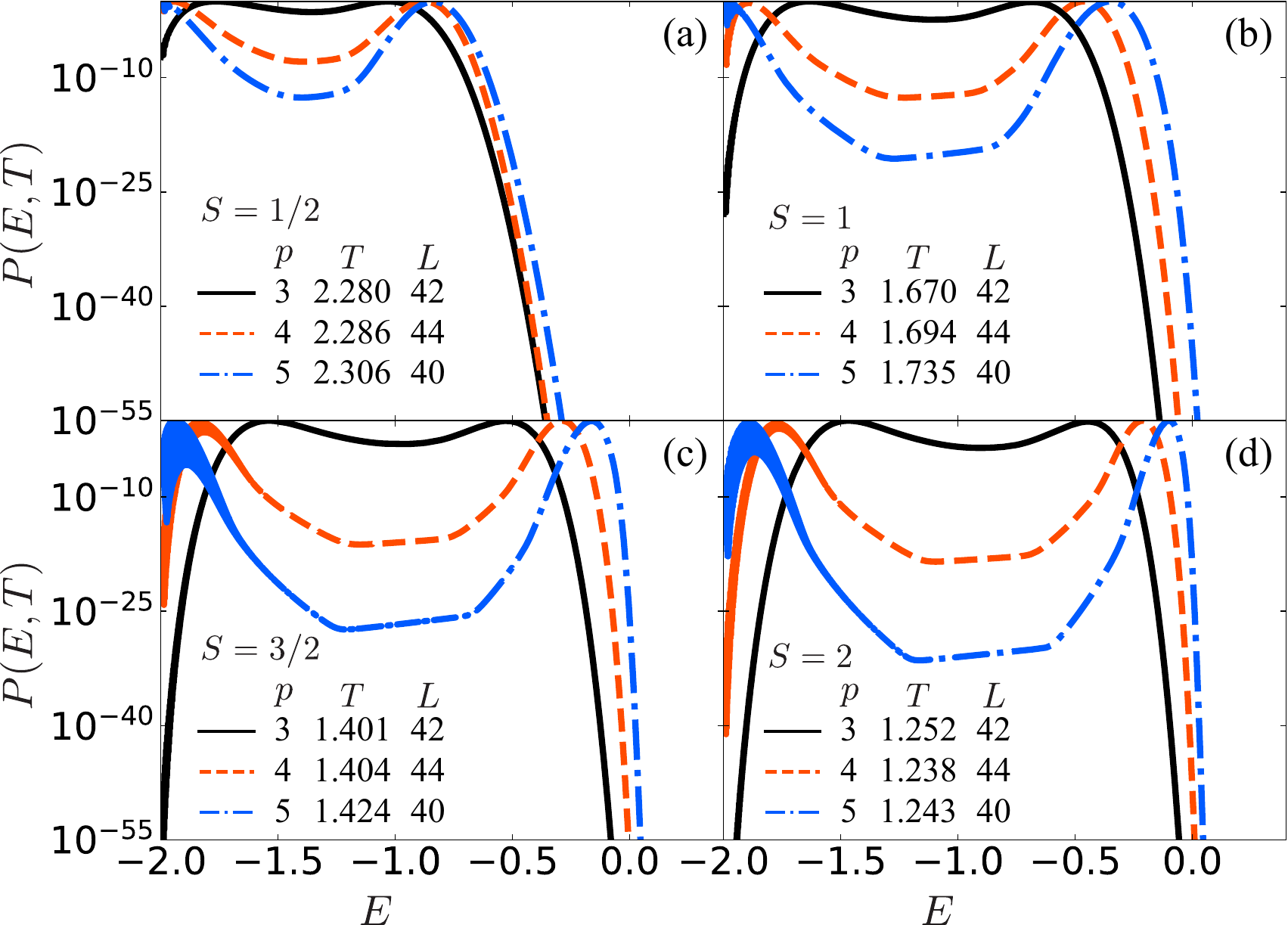}
  \caption{
    Energy $E$ dependence of canonical energy distribution $P(E,T)$ near the transition temperature for (a) $S=1/2$, (b) $S=1$, (c) $S=3/2$, and (d) $S=2$.
    The distribution is normalized for its maximum value to be 1.
    The data are obtained from single multicanonical simulation.
  }
  \label{2_dim_energy_dist}
\end{figure}

To more directly confirm that the system undergoes a first-order transition, we examine the canonical energy distribution $P(E,T)$, which is defined as follows:
\begin{align}
    P(E,T)\coloneqq D(E)e^{-\beta E}.
\end{align}
The energy dependence of $P(E, T)$ near the transition temperature is shown in Fig. \ref{2_dim_energy_dist}.
Note that $P(E, T)$ is normalized for its maximum value to be 1.
For all values of $p$ and $S$, the energy distributions exhibit double-peak structures, showing a characteristic feature of first-order phase transitions \cite{PhysRevB.34.1841}. 
Furthermore, the energy gap between two peaks widens as $p$ and $S$ increase, reflecting an increase in latent heat for first-order phase transitions. 
These results are consistent with those of the order parameter and the internal energy.

\begin{figure}[!t]
  \centering
  \includegraphics[width=\columnwidth]{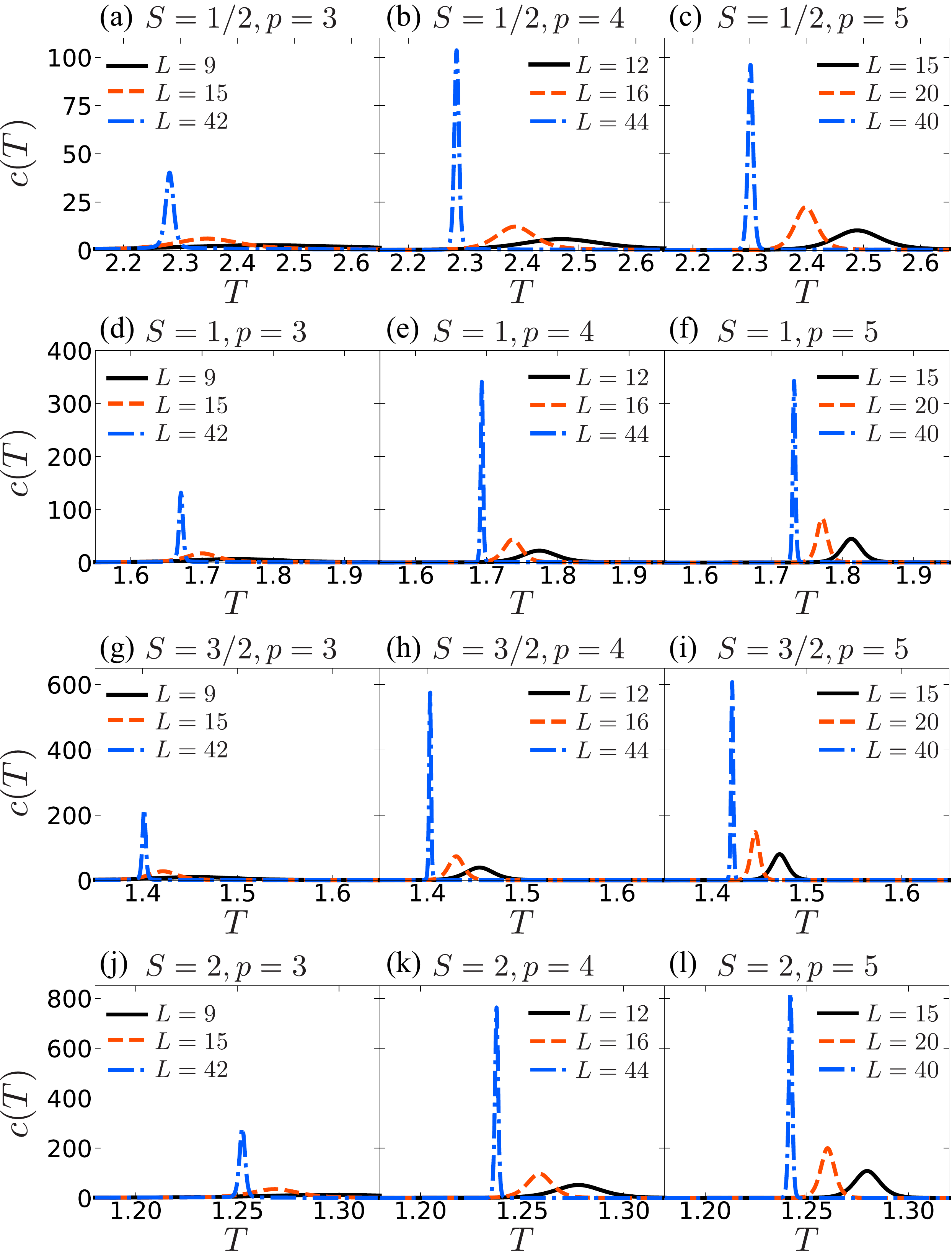}
  \caption{
    Temperature $T$ dependence of the specific heat per system size $c(T)$ for
    (a) $S=1/2,p=3$, (b) $S=1/2,p=4$, (c) $S=1/2,p=5$, (d) $S=1,p=3$, (e) $S=1,p=4$, (f) $S=1,p=5$,
    (g) $S=3/2,p=3$, (h) $S=3/2,p=4$, (i) $S=3/2,p=5$, (j) $S=2,p=3$, (k) $S=2,p=4$, and (l) $S=2,p=5$.
    Average values from five independent runs are shown here.
    The maximum standard deviation is $\sim 10^{-3}$ and too small to be visible on the graph for all results.
  }
  \label{2_dim_specific_heat}
\end{figure}

\begin{figure}[!t]
  \centering
  \includegraphics[width=\columnwidth]{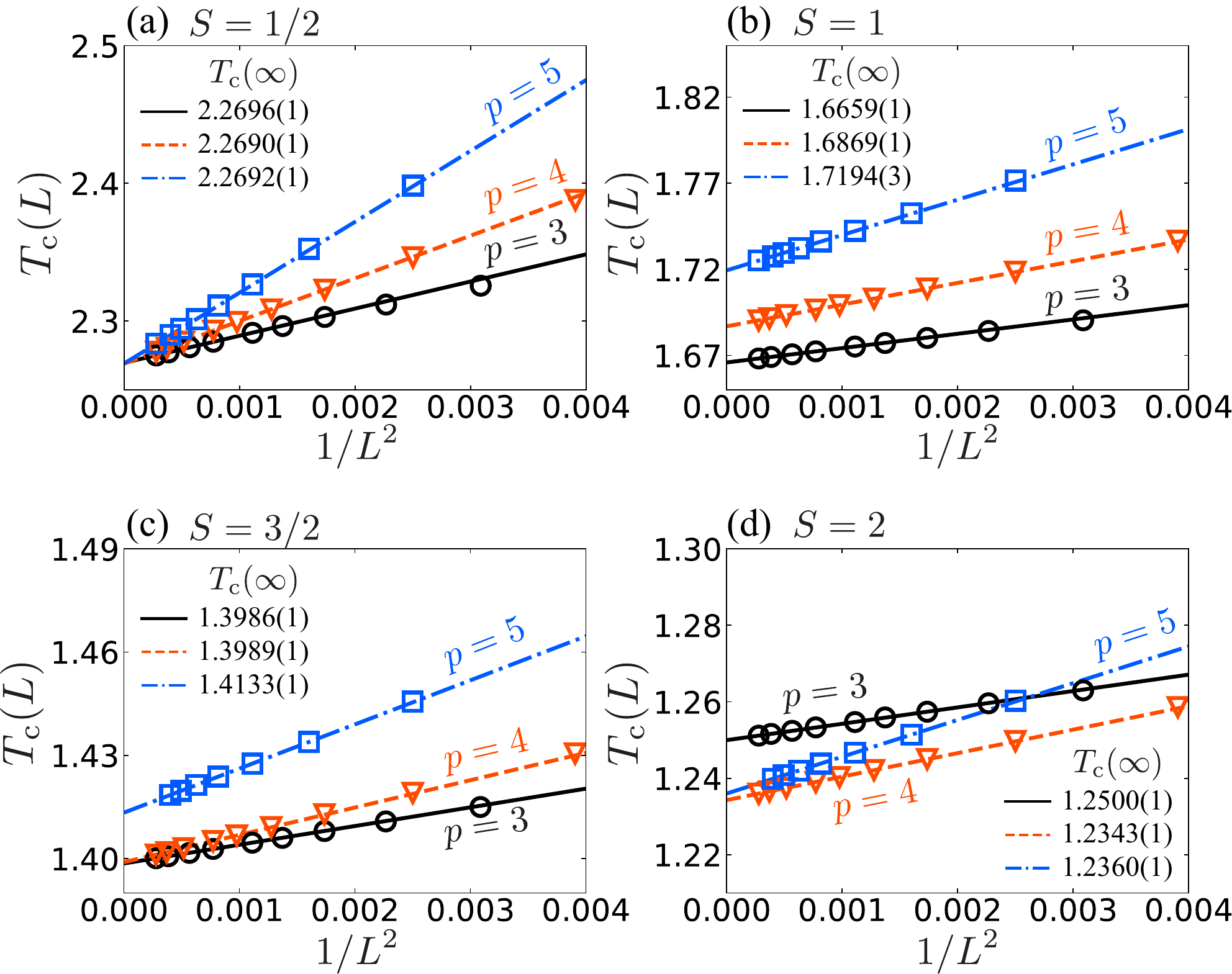}
  \caption{
    Inverse system size dependence of finite-size ``transition temperatures" $T_{\rm c}(L)$ for (a) $S=1/2$, (b) $S=1$, (c) $S=3/2$, and (d) $S=2$.
    Average values from five independent runs are shown here.
    The maximum standard deviation is $\sim 10^{-4}$ and too small to be visible in the plotted data.
    The solid and dashed lines represent the fits by $T_{\rm c}(L) = a/L^2 + T_{\rm c}(\infty)$.
    Fitting was performed using five data points in the range of $L=30$ to 60.
  }
  \label{2_dim_Tc}
\end{figure}

So far, we have discussed the finite-size results. 
We then estimate physical quantities in the thermodynamic limit from these results.
We start by discussing the transition temperature.
For this purpose, we define finite-size ``transition temperature" as the temperature which gives the maximum value of the specific heat:
\begin{align}
    T_{\text{c}}(L)\coloneqq \arg \max_{T} c(T).
    \label{TcL}
\end{align}
Here, the specific heat is defined as the derivative of the internal energy density and can be calculated by
\begin{align}
    c(T)\coloneqq\frac{\partial}{\partial T}u(T)=\frac{1}{L^2}\frac{\braket{E^2} - \braket{E}^2}{T^2}.
\end{align}
In finite-size numerical calculations for a system undergoing a first-order transition, the specific heat tends to diverge as the system size $L$ increases near the transition temperature because the internal energy tends to jump but does not do so exactly at the transition point.
The temperature dependence of the specific heat is shown in Fig. \ref{2_dim_specific_heat}.
One can see that $c(T)$ actually tends to diverge near the transition temperature as the system size increases.
$T_{\text{c}}(L)$ defined in Eq. (\ref{TcL}) represents the temperature at which this divergence tendency is observed.
This quantity behaves as follows with increasing system size:
\begin{align}
    T_{\text{c}}(L) = \frac{a}{L^2} + T_{\text{c}}(\infty).
    \label{Tc_scaling}
\end{align}
Here, $a$ is a constant value and $T_{\text{c}}(\infty)$ is the estimated transition temperature in the thermodynamic limit.

The inverse system size $1/L^2$ dependence of $T_{\text{c}}(L)$ is shown in Fig. \ref{2_dim_Tc}.
The solid and dashed lines represent the fits by Eq. (\ref{Tc_scaling}).
It can be seen that $T_{\rm c}(L)$ follows the form given in Eq. (\ref{Tc_scaling}) for larger $L$ and the fitting errors are very small.
For $S=1/2$, the transition temperatures for $p=3,4,5$ take almost the same values: $T_{\rm c}(\infty)\simeq 2.269$.
This value is very close to $2/\log(\sqrt{2} + 1)=2.269185...$, which is the exact transition temperature for $p=2$.
The fact that the transition temperature for $S=1/2$ takes the same value for all $p$ is consistent with the analytical results obtained using self-duality \cite{LTurban_1982}.
In contrast, for $S=1$, the transition temperatures for $p=3,4,5$ are different and increase with increasing $p$.
Interestingly, for $S=3/2$, the transition temperatures for $p=3$ and 4 take almost the same values: $T_{\text{c}}(\infty) \simeq 1.398$.
From the finite-size numerical calculations, it is, however, difficult to determine whether this close similarity is merely coincidental or if the transition temperatures are exactly the same for some theoretical reasons.
The case of $S=2$ shows no simple correlation between transition temperatures and $p$.

\begin{figure}[!t]
  \centering
  \includegraphics[width=\columnwidth]{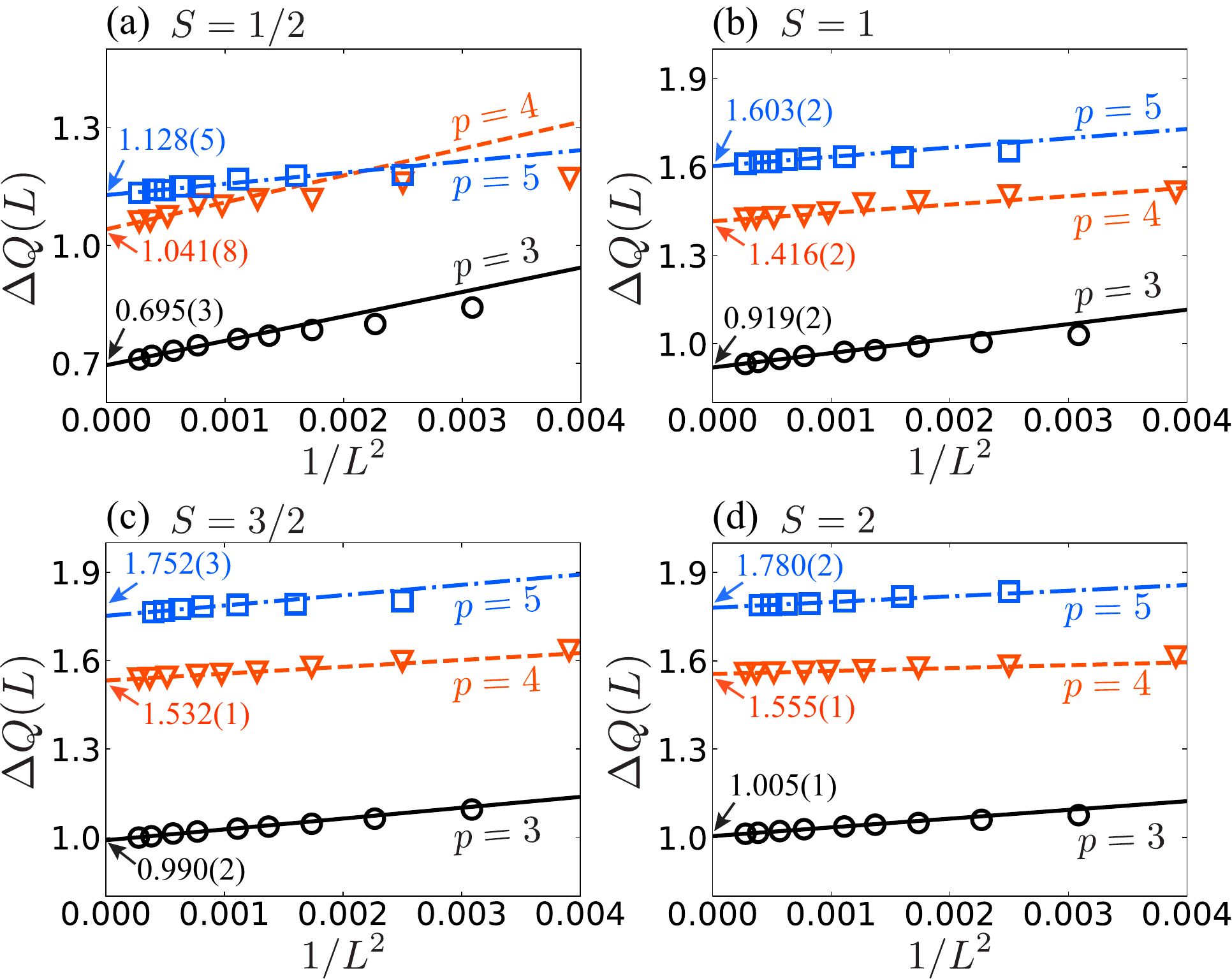}
  \caption{
    Inverse system size dependence of the latent heat $\Delta Q(L)$ for (a) $S=1/2$, (b) $S=1$, (c) $S=3/2$, and (d) $S=2$.
    Average values from five independent runs are shown here.
    The maximum standard deviation is $\sim 10^{-3}$ and too small to be visible in the plotted data.
    The solid and dashed lines represent the fits by $\Delta Q(L) = a/L^2 + \Delta Q(\infty)$.
    Fitting was performed using five data points in the range of $L=30$ to 60.
  }
  \label{2_dim_latent_heat}
\end{figure}

Then we discuss the latent heat in the thermodynamic limit characteristic of first-order phase transitions.
We denote the energies that correspond to the two peaks of $P(E, T_{\text{c}}(L))$ as $E_{-}(L)$ and $E_{+}(L)$.
Note that $E_{-}(L) < E_{+}(L)$.
Using both energies, we define finite size ``latent heat" as 
\begin{align}
    \Delta Q(L)\coloneqq E_{+}(L) - E_{-}(L).
\end{align}
Similar to $T_{\text{c}}(L)$, this quantity behaves as follows with increasing system size:
\begin{align}
    \Delta Q(L) = \frac{a}{L^2} + \Delta Q(\infty).
    \label{Q_scaling}
\end{align}
Here, $a$ is a constant and $\Delta Q(\infty)$ can be considered as the latent heat in the thermodynamic limit.
Figure \ref{2_dim_latent_heat} shows the inverse system size dependence of $\Delta Q(L)$.
One can see that $\Delta Q(L)$ behaves as Eq. (\ref{Q_scaling}) for large $L$ and the fitting errors are small.
The overall trend is that the latent heat increases with $p$ for each $S$. 
This indicates that the system exhibits a stronger first-order transition as $p$ increases. It is also consistent with the results of the order parameter and internal energy.

\begin{figure}[!t]
  \centering
  \includegraphics[width=\columnwidth]{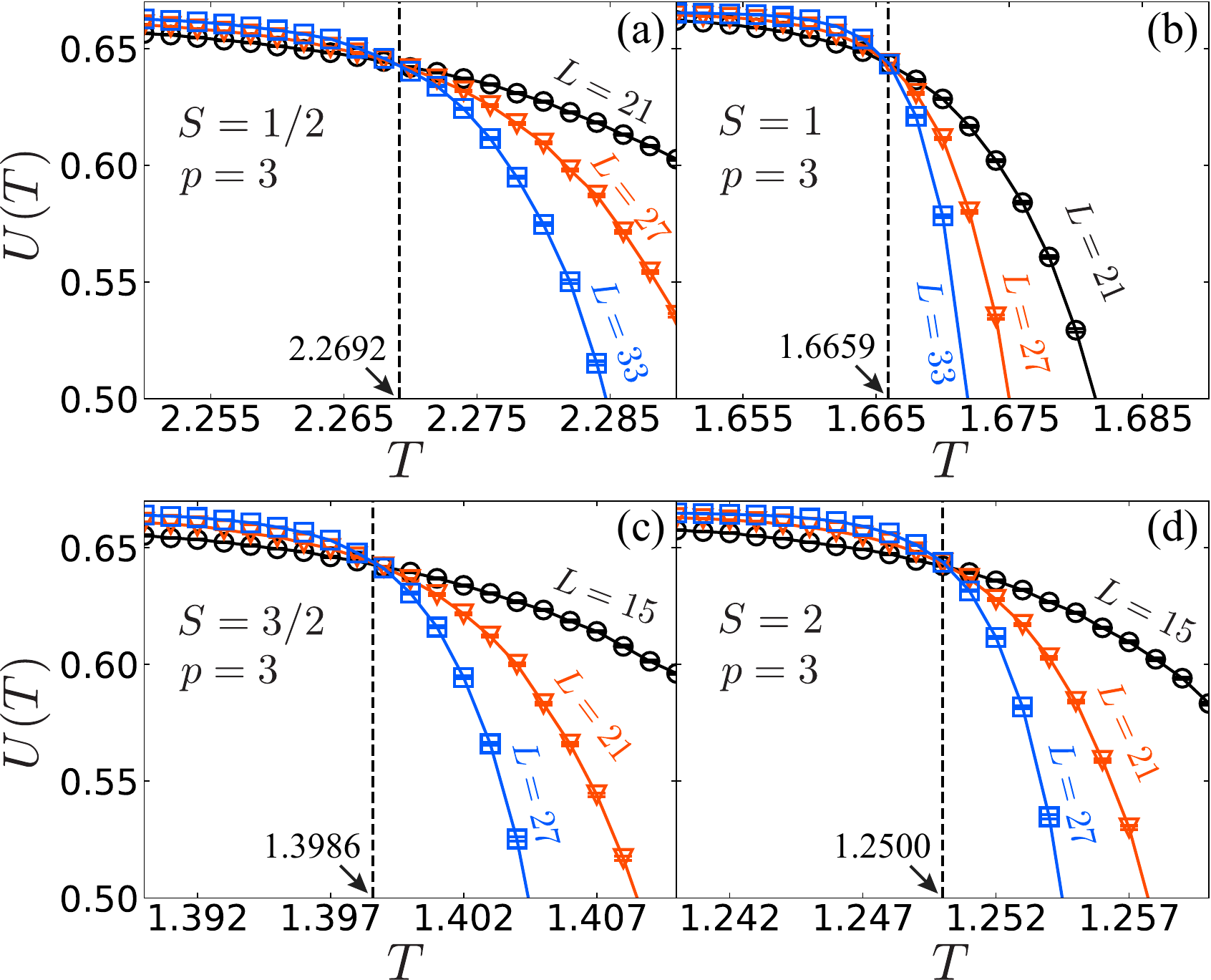}
  \caption{
    Temperature $T$ dependence of the Binder cumulant $U(T)$ near the transition temperature for $p=3$ and (a) $S=1/2$, (b) $S=1$, (c) $S=3/2$, and (d) $S=2$.
    The dashed vertical lines represent the estimated transition points obtained from Eq. (\ref{Tc_scaling}).
  }
  \label{2_dim_binder}
\end{figure}

To confirm that the systems show first-order transitions at the temperatures obtained here, we calculate the following Binder cumulant:
\begin{align}
    U(T)\coloneqq 1 - \frac{\braket{O^2}}{3\braket{O}^2}.
\end{align}
It is known that the Binder cumulants for different system sizes intersect at the transition temperature for sufficiently large system sizes
\cite{PhysRevB.30.1477,doi:10.1142/S0129183192000683,doi:10.1080/00150191003670291}.
We here employed a Monte Carlo simulation with single-spin flips to perform the calculations, ensuring consistency in our results obtained by the multicanonical simulations.
This method requires a large number of sweeps to converge and is difficult to carry out for the models for $p \geq 4$, whereas the $p=3$ models converge with a relatively small number of sweeps $\sim 10^4$.
Figure \ref{2_dim_binder} shows $U(T)$ near the transition temperature for $p=3$ and $S=1/2,1,3/2,2$.
The Metropolis update \cite{METROPOLIS,METROPOLIS_HASTING} is used for $S=1/2$ and the Suwa-Todo method \cite{PhysRevLett.105.120603} for $S \geq 1$.
The number of sweeps is up to $5\times 10^4$ and the number of samples is $10^5$ for all data.
It is evident that $U(T)$ intersects at the estimated transition temperature for all $S$, which is consistent with the results obtained by the multicanonical simulations.

\section{Summary}\label{sect-summary}
We have analyzed the spin-$S$ Ising models with $p$-spin interactions on the one-dimensional chain and the two-dimensional square lattice, varying both $S$ and $p$ up to $S=2$ and $p=5$ to explore how these changes affect the properties of the system.
For the one-dimensional model, we have calculated the free energy and the spin-spin correlations by numerically diagonalizing the transfer matrix. 
It is found that the free energy decreases as $p$ increases for $S \geq 1$, while the free energy is independent of $p$ for $S=1/2$.
Additionally, the spin-spin correlations increase with $p$ and this increase becomes significant in the low-temperature region.
The correlation length $\xi_p$ takes the form $\xi_p\simeq Sp\times\xi_{p=2}$ for $p\geq 3$ near zero temperature.

\begin{table}[t!]
  \caption{
    The transition temperatures $T_{\text{c}}(\infty)$ and the latent heat $\Delta Q(\infty)$ estimated from the scaling forms in Eqs. (\ref{Tc_scaling}) and (\ref{Q_scaling}), respectively.
    For comparison, the values for $p=2$ are also shown.
  }
  \label{table_Tc_heat}
  \centering
  \begin{tabular}{ccccc}
    \hline\hline
    $S$ & $p$ & $T_{\text{c}}(\infty)$ & $\Delta Q(\infty)$ & Transition order \\
    \hline
    1/2 & 2 & 2.26918... \cite{PhysRev.65.117} &  & Second \\
        & 3 & 2.2696(1) & 0.695(3) & First \\
        & 4 & 2.2690(1) & 1.041(8) & First \\
        & 5 & 2.2692(1) & 1.128(5) & First \\
    \hline
    1   & 2 & 1.68--1.71 \cite{PhysRevB.33.1717,PhysRevB.57.11575,PhysRevE.73.036702,PhysRevE.81.041113,BUTERA201822} &  & Second \\
        & 3 & 1.6659(1) & 0.919(2) & First \\
        & 4 & 1.6869(1) & 1.416(2) & First \\
        & 5 & 1.7194(3) & 1.603(2) & First \\
    \hline
    3/2 & 2 & 1.46 \cite{PhysRevB.57.11575} &  & Second \\
        & 3 & 1.3986(1) & 0.990(2) & First \\
        & 4 & 1.3989(1) & 1.532(1) & First \\
        & 5 & 1.4133(1) & 1.752(3) & First \\
    \hline
    2   & 2 & 1.32--1.69 \cite{TUCKER198727,KANEYOSHI1992495,JURCISIN1996684,HACHEM2017927} &  & Second \\
        & 3 & 1.2500(1) & 1.005(1) & First \\
        & 4 & 1.2343(1) & 1.555(1) & First \\
        & 5 & 1.2360(1) & 1.780(2) & First \\
    \hline\hline
  \end{tabular}
\end{table}

For the model on the two-dimensional square lattice, we have investigated the nature of the phase transition.
By using the multicanonical methods, we have analyzed basic physical quantities such as the order parameter, internal energy, and heat capacity.
It is found that a first-order phase transition occurs for $p \geq 3$. 
While it is well known that $S = 1/2$ Ising-like models with higher-order interactions exhibit first-order transitions \cite{MDebierre_1983, BLOTE1986395, PhysRevLett.63.1546, pssb.2221750217, NotesPSPINS, PhysRevE.88.012140, JURCISINOVA2014375}, 
we have shown that this also holds true for $S \geq 1$.
Furthermore, analysis of the latent heat shows that increasing $p$ or $S$ leads to a stronger first-order transition, with a corresponding increase in the latent heat.
Regarding the transition temperature, our results suggest that for $S=1/2$, the transition temperature is independent of $p$ and is close to $T_{\text{c}}=2/\log(\sqrt{2}+1)$, as previously predicted using self-duality \cite{LTurban_1982}.
However, for $S \geq 1$, the relation between $p$ and the transition temperature is not simple. 
When $p$ is fixed and $S$ increases, the transition temperature tends to decrease, and it is expected to converge to a finite value in the limit of infinite $S\rightarrow\infty$.
For $S=3/2$, the transition temperatures for $p=3$ and $4$ are very close, with $T_{\text{c}} \simeq 1.398$. 
Whether this close similarity is coincidental or has a theoretical background remains an open question for future studies.
We summarize the transition temperatures and the latent heat obtained in this paper in Table \ref{table_Tc_heat}, along with the results for $p=2$ for comparison.
The results obtained in this paper are expected to deepen our understanding of the properties of $p$-body interactions and serve as a foundation for various future studies, including potential applications.

Finally, let us comment on future works.
One interesting point is exploring Ising models with continuous spins, obtained by $S\rightarrow\infty$.
While the transition temperature decreases as the magnitude of spin $S$ increases for fixed $p$, the latent heat tends to increase, suggesting that the first-order transition could become stronger.
In addition to the $p$-spin interaction $J_p$ considered in this paper, introducing conventional two-body interactions $J_2$ is also interesting.
It is expected that a system with sufficiently large $J_2$ shows a second-order phase transition, 
while for sufficiently small $J_2$, the $p$-spin interactions become dominant and the system shows a first-order transition.
As $J_2$ decreases from large values, there could appear a tricritical point, where the phase transition changes from second order to first order. 
Analyzing this would also be an interesting topic for future research.

\section*{Acknowledgement}
We would like to thank research members of Jij Inc. for their helpful discussion and conversation.

\bibliography{ref}
\end{document}